\documentclass[prd,showpacs,twocolumn]{revtex4}

\usepackage{amsmath}
\usepackage{latexsym}
\usepackage{graphicx}
\usepackage{epstopdf}
\usepackage{mathbbol}
\usepackage{dsfont}

\begin{document}


\title{Using curvature invariants for wave extraction in numerical relativity} 


\author{Andrea~Nerozzi}
\affiliation{
Institut f\"ur Angewandte Mathematik,
Friedrich-Schiller-Universit\"at,
Ernst-Abbe-Platz 2, 07743 Jena, Germany
}

\author{Oliver~Elbracht}
\affiliation{
Institut f\"ur Theoretische Physik und Astrophysik,
Julius-Maximilians-Universit\"at,
Am Hubland, 97074 W\"urzburg, Germany
}


\date{\today}


\begin{abstract}
We present a new expression for the Weyl scalar $\Psi_4$ that can be used in numerical
relativity to extract the gravitational wave content of a spacetime. The formula relies
upon the identification of transverse tetrads, namely the ones in which $\Psi_1=\Psi_3=0$. It is well
known that tetrads with this property always exist in a general Petrov type I 
spacetime. A sub-class of these tetrads naturally converges to the Kinnersley tetrad 
in the limit of Petrov type D spacetime. However, the transverse condition fixes only four
of the six parameters coming from the Lorentz group of transformations applied to tetrads.
Here we fix the tetrad completely, in
particular by giving the expression for the spin-boost transformation that was still
unclear. The value of $\Psi_4$ in this optimal tetrad is given as a function of the
two curvature invariants $I$ and $J$.
\end{abstract}




\maketitle

\section{Introduction}
\label{sec:intro}

With the recent breakthroughs in numerical relativity 
\cite{Pretorius:2007nq,pretorius-2005-95, baker-2006-96, baker-2006-73, Campanelli:2005dd, Sperhake:2006cy,Pollney:2007ss,Boyle:2007ft,Herrmann:2007ac,
Sperhake:2008ga,Gonzalez:2007hi,Gonzalez:2006md,Campanelli:2007cga,
Campanelli:2007ew,Rezzolla:2007rd,Brugmann:2008zz,Buonanno:2006ui,Boyle:2008ge,
Berti:2007fi,Healy:2008js}
concerning the numerical
evolution of binary black hole systems, the problem of extracting waveforms
from a numerically evolved spacetime has become of primary importance. One
of the most used techniques for wave extraction involves the
Newman-Penrose formalism and in particular the calculation of the
Weyl scalar $\Psi_4$. 

The Newman-Penrose formalism is a tetrad formalism where the
tetrad vectors are chosen to be null, two of them being real
(normally referred to as $\ell^{\mu}$ and $n^{\mu}$),
the other two complex conjugated (normally referred to as
$m^{\mu}$ and $\bar{m}^{\mu}$). The idea underlying this approach 
is that by contracting the Weyl tensor on null vectors, the physical
properties of the spacetime are easier to single out. 

The relevant quantities in this formalism are the Weyl scalars, given by the
contraction of the Weyl tensor over a specific combination of the four null tetrad 
vectors, and the connection coefficients (spin
coefficients) related to the covariant derivatives of the same tetrad vectors.
All of these quantities are scalars, making the choice of the coordinate system
irrelevant for their calculation; they are however dependent on the tetrad
choice which constitutes the gauge freedom in this formalism.

The normal equations governing the gravitational
field, namely the Bianchi and Ricci identities, can be rewritten in this formalism as
functions of the Weyl scalars and the spin coefficients, 
and a detailed presentation
can be found in \cite{Chandrasekhar83}. 
However, the whole set of equations is clearly higher in number than the
real physical degrees of freedom, and it is not clear how these equations are related to each 
other. 

On the other hand, it is well known that the variables introduced within this formalism, under certain assumptions, acquire a precise
physical meaning. For example $\Psi_0$ and $\Psi_4$ are related
to the ingoing and outgoing gravitational wave contribution, while $\Psi_2$ is related to the background
contribution to the curvature.

The spin coefficients can also be related to physical properties of the tetrad vectors: 
$k$ and $\epsilon$ are related to the geodesic properties of the $\ell^{\mu}$ vector, $\rho$
plays a determinant role in establishing whether $\ell^{\mu}$ is hypersurface orthogonal or not, and
$\sigma$ measures the shear of null geodesics defined by $\ell^{\mu}$.

For the mentioned physical properties to hold, a suitable tetrad choice is fundamental. 
The importance of a robust wave extraction technique has been underlined by some
recent articles \cite{Lehner:2007ip,Gallo:2008pm,Gallo:2008sk,Pazos:2006kz}.
Recent works \cite{beetle-2005-72, nerozzi-2005-72, nerozzi-2006-73,nerozzi-2006-,
nerozzi-2007-75,campanelli-2006-73b}
have identified in 
{\it transverse tetrads}, i.e. those tetrads satisfying the condition $\Psi_1=\Psi_3=0$,
a convenient candidate for wave extraction and in general for a better understanding of
the equations governing the Newman-Penrose formalism. 
Transverse tetrads can always be found in a general Petrov type I spacetime: we know
in fact \cite{beetle-2002-89} that there are three families (frames) of transverse tetrads. The degeneracy in
each single frame is due to the fact that the condition $\Psi_1=\Psi_3=0$ does not fix
the tetrad 
completely, leaving the spin-boost (type III rotation) degree
of freedom yet to be specified. Therefore all the tetrads in every transverse frame
can be related by a spin-boost transformation. 

The advantages of the choice $\Psi_1=\Psi_3=0$ have already been shown in
\cite{beetle-2005-72, nerozzi-2005-72, nerozzi-2006-73,nerozzi-2006-,
nerozzi-2007-75,campanelli-2006-73b}, namely one of these three frames naturally converges to the
frame where $\Psi_0=\Psi_4=0$ when the spacetime approaches Petrov type D. This property ensures that the values for the two
scalars $\Psi_0$ and $\Psi_4$, in the linear regime, are at first order tetrad invariant, and
directly associated with the gravitational wave signal \cite{teukolsky73}.

To determine the tetrad completely, however, one also needs to fix the spin-boost
degree of freedom. The Kinnersley tetrad \cite{Kinnersley69} 
identifies the spin-boost parameter by imposing the condition 
$\epsilon=0$. The motivation for this choice is related
to the physical properties of the $\ell$ null vector:  the geodesic equation
for $\ell$ reads

\begin{equation}
\ell^{\mu}\nabla_{\mu}\ell^{\nu}=\left(\epsilon+\epsilon^*\right)\ell^{\nu}-\kappa \bar{m}^{\nu}-
\kappa^* m^{\nu},
\label{eqn:geodl}
\end{equation}
which shows that if the two spin coefficients $\kappa$ and $\epsilon$ vanish, the vector $\ell^{\mu}$
is geodesic and affinely parametrized. In the limit of Kerr spacetime, the Goldberg-Sachs theorem
\cite{Chandrasekhar83} 
guarantees that $\kappa=0$, so the additional condition $\epsilon=0$ enforces the
affine parametrization of $\ell^{\mu}$.

In order to find transverse frames in a numerical simulation, one procedure is to calculate
the Weyl scalars using an initial tetrad, and then calculate the rotation parameters for type
I and type II rotations using the two methods given in \cite{beetle-2005-72, nerozzi-2005-72}. This has been applied
successfully in \cite{campanelli-2006-73b}. This procedure is rather lengthy to apply in practice;
moreover, it does not fix the spin-boost parameter in a rigorous way. Here,
we want to validate a different and simplified approach that can be used when
one is not interested in the expression of the tetrad vectors, but only in the final expression
for $\Psi_4$ in the right tetrad. This procedure gives a rigorous expression for the spin-boost
parameter by enforcing the condition $\epsilon=0$ in the Petrov type D limit, and the
final result for the scalars $\Psi_0$, $\Psi_2$, and $\Psi_4$ is given as functions of the
two curvature invariants $I$ and $J$. 

The paper is organized as follows: Sec.~\ref{eqn:general} presents some general definitions of the
relevant quantities in the Newman-Penrose formalism and provides an equation
for the three Weyl scalars $\Psi_0$, $\Psi_2$, and $\Psi_4$ that are valid in transverse
frames where $\Psi_1=\Psi_3=0$. Sec.~\ref{sec:bianchi} analyzes the Bianchi identities
in the limit of Petrov type D spacetime in order to obtain information on the spin coefficient
$\epsilon$ and its connection to the spin-boost parameter. We will show that the Bianchi
identities provide only information on other spin coefficients in the limit of Petrov
type D spacetime. An expression for $\epsilon$ is then 
obtained using the Ricci identities in Sec.~\ref{sec:ricci}. Finally in Sec.~\ref{sec:kerr} we 
apply the result to the case of the Kinnersley tetrad by enforcing the condition $\epsilon=0$
and obtain the corresponding spin-boost parameter value. This result
leads to the final expression for the Weyl scalars in this particular tetrad.

\section{General definitions}
\label{eqn:general}

Weyl scalars are given by contraction of the Weyl tensor over a certain combination
of four null vectors, two real ($\ell^{\mu}$ and $n^{\mu}$) and two complex conjugates
($m^{\mu}$ and $\bar{m}^{\mu}$), according to

\begin{subequations}
\label{eqn:weylscalars}
\begin{eqnarray}
\Psi_0 &=&-C_{abcd}\ell^am^b\ell^cm^d, \label{eqn:weylscalars0} \\
\Psi_1 &=&-C_{abcd}\ell^an^b\ell^cm^d, \label{eqn:weylscalars1} \\
\Psi_2 &=&-C_{abcd}\ell^am^b\bar{m}^cn^d, \label{eqn:weylscalars2} \\
\Psi_3 &=& -C_{abcd}\ell^an^b\bar{m}^cn^d, \label{eqn:weylscalars3} \\
\Psi_4 &=& -C_{abcd}n^a\bar{m}^bn^c\bar{m}^d. \label{eqn:weylscalars4} 
\end{eqnarray}
\end{subequations}

The tetrad choice constitutes the gauge degree of freedom in the calculation of Weyl
scalars, and can be represented by the six parameter Lorentz group of gauge
transformations. Despite Weyl scalars being tetrad dependent, it is possible to
construct two quantities which are no longer dependent on the tetrad choice. Such
quantities are the curvature invariants $I$ and $J$, and their
expressions as functions of the Weyl scalars are given by

\begin{subequations}
\label{eqn:curvatureinvariantsWeyl}
\begin{eqnarray}
I &=& \Psi_4\Psi_0-4\Psi_1\Psi_3+3\Psi_2^2, \label{eqn:IinvariantWeyl} \\
J &=& \textrm{det}\left|\begin{array}{ccc}
\Psi_4 & \Psi_3 & \Psi_2 \\
\Psi_3 & \Psi_2 & \Psi_1 \\
\Psi_2 & \Psi_1 & \Psi_0 
\end{array} \right|.
\label{eqn:JinvariantWeyl} 
\end{eqnarray}
\end{subequations}

If we assume to fix all the six degrees of freedom related to the tetrad choice
by requiring that $\Psi_1=\Psi_3=0$ and $\Psi_0=\Psi_4$, the expressions for the 
two curvature invariants written in Eq.~(\ref{eqn:curvatureinvariantsWeyl}) simplify
to

\begin{subequations}
\label{eqn:curvatureinvariantsWeyl2}
\begin{eqnarray}
I &=& \Psi_4^2+3\Psi_2^2, \label{eqn:IinvariantWeyl2} \\
J &=& \Psi_4^2\Psi_2-\Psi_2^3 .
\label{eqn:JinvariantWeyl2} 
\end{eqnarray}
\end{subequations}
Eq.~(\ref{eqn:IinvariantWeyl2}) and (\ref{eqn:JinvariantWeyl2})
can now be inverted to give $\Psi_2$ and $\Psi_4$ as functions of the curvature
invariants $I$ and $J$. In particular one can solve for $\Psi_2$ by setting $\Psi_2=\frac{\lambda}{2}$,
where $\lambda$ is the solution of the characteristic polynomial 

\begin{equation}
\lambda^3-2I\lambda+2J=0.
\label{eqn:polynomial}
\end{equation}
The three possible solutions are given by

\begin{subequations}
\label{eqn:solnewmodpoly}
\begin{eqnarray}
\lambda_1&=&-\left(P+\frac{I}{3P}\right), \label{eqn:sollambda1} \\
\lambda_2&=&-\left(e^{\frac{2\pi i}{3}}P+e^{\frac{4\pi i}{3}}\frac{I}{3P}\right), \label{eqn:sollambda2} \\
\lambda_3&=&-\left(e^{\frac{4\pi i}{3}}P+e^{\frac{2\pi i}{3}}\frac{I}{3P}\right), \label{eqn:sollambda3} 
\end{eqnarray}
\end{subequations}
where $P$ is defined as

\begin{equation}
P=\left[J+\sqrt{J^2-\left(I/3\right)^3}\right]^{\frac{1}{3}}.
\label{eqn:pdefinition}
\end{equation}
Setting $\Psi_2$ equal to the three possible roots gives the three possible transverse
frames. $\Psi_4$ is then determined accordingly once we have fixed a specific 
transverse frame.

Setting $\Psi_0=\Psi_4$ for the spin-boost degree of freedom
is not the best possible choice, 
since in this case,  the two Weyl scalars 
have the radial fall-off of $r^{-3}$ at future null infinity. We can nevertheless use this choice as a starting point 
and reinsert the spin-boost degree of freedom into the expressions for the scalars:
introducing the quantity
$\Theta=\sqrt{3}P I^{-\frac{1}{2}}$ 
and the spin-boost
parameter $\mathcal{B}=\left(\frac{\Psi_4}{\Psi_0}\right)^{\frac{1}{4}}$, the three non
vanishing Weyl scalars can be written as

\begin{subequations}
\label{eqn:psi2psi4qkt}
\begin{eqnarray}
\Psi_0&=&-\frac{i\mathcal{B}^{-2}}{2}\cdot \Psi_{-},
\label{eqn:psi0qkt} \\
\Psi_2&=&-\frac{1}{2\sqrt{3}}\cdot\Psi_{+},
\label{eqn:psi2qkt} \\
\Psi_4&=&-\frac{i\mathcal{B}^{2}}{2}\cdot \Psi_{-},
\label{eqn:psi4qkt} 
\end{eqnarray}
\end{subequations}
where

\begin{equation}
\Psi_{\pm}= I^{\frac{1}{2}}\left(e^{\frac{2\pi i k}{3}}\Theta\pm e^{-\frac{2\pi i k}{3}}\Theta^{-1}\right),
\label{eqn:pispm}
\end{equation}
and $k$ is an integer number assuming 
the values $\left\{0,1,2\right\}$ corresponding to the three different transverse frames.
For the principal root in the expression for $P$ given in 
Eq.~(\ref{eqn:pdefinition}), the limit of type D corresponds to $\Theta\rightarrow 1$. Subsequently,
from Eq.~(\ref{eqn:pispm}), the frame with $k=0$ is the frame where
$\Psi_-$ (and consequently  $\Psi_0$ and $\Psi_4$) tends to zero, i.e. the transverse frame
which is also a quasi-Kinnersley frame \cite{nerozzi-2005-72}.

This paper demonstrates how the expressions in Eq. (\ref{eqn:psi2psi4qkt}-\ref{eqn:pispm}) for the Weyl
scalars relate to the
expressions for the spin coefficients in the limit of Petrov type D spacetime; 
we obtain an expression for the spin coefficient
$\epsilon$ and enforce the condition $\epsilon=0$. To do this, we introduce 
the
directional derivative operators along the null tetrad vectors
$D=\ell^{\mu}\nabla_{\mu}$,
$\Delta=n^{\mu}\nabla_{\mu}$, $\delta=m^{\mu}\nabla_{\mu}$ and 
$\delta^*=\bar{m}^{\mu}\nabla_{\mu}$, and analyze the Bianchi and Ricci identities 
in the Newman-Penrose formalism.

\section{The Bianchi identities}
\label{sec:bianchi}

Setting $\Psi_1=\Psi_3=0$ simplifies the Bianchi identities, given here in terms of the Weyl scalars and spin coefficients

\begin{subequations}
\label{eqn:bianchiidlist}
\begin{eqnarray}
D\Psi_4&=&-3\lambda\Psi_2-\left(4\epsilon-\rho\right)\Psi_4,
\label{eqn:bianchiD4} \\
D\Psi_2&=&-\lambda\Psi_0+3\rho\Psi_2,
\label{eqn:bianchiD2} \\
\Delta\Psi_0&=&3\sigma\Psi_2+\left(4\gamma-\mu\right)\Psi_0,
\label{eqn:bianchiDelta4} \\
\Delta\Psi_2&=&\sigma\Psi_4-3\mu\Psi_2,
\label{eqn:bianchiDelta2} \\
\delta\Psi_4&=&-3\nu\Psi_2+\left(\tau-4\beta\right)\Psi_4,
\label{eqn:bianchidelta4} \\
\delta\Psi_2&=&-\nu\Psi_0+3\tau\Psi_2,
\label{eqn:bianchidelta2} \\
\delta^*\Psi_0&=&3\kappa\Psi_2+\left(4\alpha-\pi\right)\Psi_0,
\label{eqn:bianchideltastar4} \\
\delta^*\Psi_2&=&\kappa\Psi_4-3\pi\Psi_2.
\label{eqn:bianchideltastar2} 
\end{eqnarray}
\end{subequations}

A key ingredient is rewriting the
Bianchi identities in terms of the newly introduced variables $\Psi_{\pm}$. 
Since $\epsilon$ appears only in the first two Bianchi identities, we will give
the details of the calculation only for the derivative operator $D$; however, as the 
symmetry of the Bianchi identities suggests, the calculation for the other derivatives
is analogous, and we will use this property at the end of this paper to
calculate the expressions for the spin coefficients $\gamma$, $\alpha$ and $\beta$.

Using Eq.~(\ref{eqn:psi2psi4qkt}) to relate the Weyl scalars $\Psi_0$, $\Psi_2$ and $\Psi_4$ to the
new scalars $\Psi_+$ and $\Psi_-$ one obtains

\begin{subequations}
\label{eqn:bianchimod}
\begin{eqnarray}
D\Psi_{+}&=&-\tilde{\lambda}\Psi_{-}+3\rho\Psi_{+},
\label{eqn:bianchimodD4} \\
D\Psi_{-}&=&\tilde{\lambda}\Psi_{+}-\left(4\tilde{\epsilon}-\rho\right)\Psi_{-},
\label{eqn:bianchimodD2} \\
\Delta\Psi_{+}&=&\tilde{\sigma}\Psi_{-}-3\mu\Psi_{+},
\label{eqn:bianchimodDelta4} \\
\Delta\Psi_{-}&=&-\tilde{\sigma}\Psi_{+}+\left(4\tilde{\gamma}-\mu\right)\Psi_{-},
\label{eqn:bianchimodDelta2} \\
\delta\Psi_{+}&=&-\tilde{\nu}\Psi_{-}+3\tau\Psi_{+},
\label{eqn:bianchimoddelta4} \\
\delta\Psi_{-}&=&\tilde{\nu}\Psi_{+}-\left(4\tilde{\beta}-\tau\right)\Psi_{-},
\label{eqn:bianchimoddelta2} \\
\delta^*\Psi_{+}&=&\tilde{\kappa}\Psi_{-}-3\pi\Psi_{+},
\label{eqn:bianchimoddeltas4} \\
\delta^*\Psi_{-}&=&-\tilde{\kappa}\Psi_{+}+\left(4\tilde{\alpha}-\pi\right)\Psi_{-},
\label{eqn:bianchimoddeltas2} 
\end{eqnarray}
\end{subequations}
where we have introduced the rescaled spin coefficients $\tilde{\lambda}=i\sqrt{3}\lambda\mathcal{B}^{-2}$,
$\tilde{\sigma}=i\sqrt{3}\sigma\mathcal{B}^{2}$, $\tilde{\nu}=i\sqrt{3}\nu\mathcal{B}^{-2}$,
$\tilde{\kappa}=i\sqrt{3}\kappa\mathcal{B}^{2}$, $\tilde{\epsilon}=\epsilon+\frac{1}{2}D\ln\mathcal{B}$,
$\tilde{\gamma}=\gamma+\frac{1}{2}\Delta\ln\mathcal{B}$, $\tilde{\beta}=\beta+\frac{1}{2}\delta\ln\mathcal{B}$, $\tilde{\alpha}=\alpha+\frac{1}{2}\delta^*\ln\mathcal{B}$.

This new set of rescaled
spin coefficients 
now transforms in the same way under
a spin-boost transformation: for example the three spin coefficients
$\left\{\rho,\tilde{\lambda},\tilde{\epsilon}\right\}$ transform as
 $\rho\rightarrow\left|\mathcal{B}\right|^{-1}\rho$,
$\tilde{\epsilon}\rightarrow\left|\mathcal{B}\right|^{-1}\tilde{\epsilon}$ and
$\tilde{\lambda}\rightarrow\left|\mathcal{B}\right|^{-1}\tilde{\lambda}$, and
analogous transformations for the other spin coefficients. This
is not surprising: $\Psi_{+}$ and $\Psi_{-}$ are only functions of curvature invariants,
so in Eq.~(\ref{eqn:bianchimodD4}) and (\ref{eqn:bianchimodD2}) the only dependence on the spin-boost parameter
on the left hand side 
comes from the $D$ derivative operator which carries a $\left|\mathcal{B}\right|^{-1}$ factor,
the right hand side must therefore be consistent and show the same spin-boost
dependence in the rescaled spin coefficients.

Dividing Eq.~(\ref{eqn:bianchimodD4}) by $\Psi_+$ and Eq.~(\ref{eqn:bianchimodD2})
by $\Psi_-$ the first two Bianchi identities become:

\begin{subequations}
\label{eqn:bianchimod2ver}
\begin{eqnarray}
\frac{D\Psi_{+}}{\Psi_+}&=&-\tilde{\lambda}\frac{\Psi_{-}}{\Psi_+}+3\rho,
\label{eqn:bianchimodD42} \\
\frac{D\Psi_{-}}{\Psi_-}&=&\tilde{\lambda}\frac{\Psi_{+}}{\Psi_-}-\left(4\tilde{\epsilon}-\rho\right).
\label{eqn:bianchimodD22} 
\end{eqnarray}
\end{subequations}

We will now study the behavior of Eq.~(\ref{eqn:bianchimod2ver}) in the Petrov type D limit.
Using Eq.~(\ref{eqn:pispm}) and applying the $D$ operator 
to $\Psi_+$ and $\Psi_-$ gives

\begin{subequations}
\label{eqn:bianchimodb}
\begin{eqnarray}
D\Psi_{+}&=&D\ln\Theta \cdot\Psi_{-}+D\ln\left(I^{\frac{1}{2}}\right)\Psi_{+},
\label{eqn:bianchimodD4b} \\
D\Psi_{-}&=&D\ln\Theta \cdot\Psi_{+}+D\ln\left(I^{\frac{1}{2}}\right)\Psi_{-}.
\label{eqn:bianchimodD2b} 
\end{eqnarray}
\end{subequations}
In the Petrov type D limit (corresponding to
$\Theta\rightarrow 1$) one has $D\Psi_{+}\rightarrow D\ln\left(I^{\frac{1}{2}}\right)\Psi_{+}$
and $D\Psi_{-}\rightarrow D\ln\left(I^{\frac{1}{2}}\right)\Psi_{-}$. This result implies that
the left hand sides in Eq.~(\ref{eqn:bianchimod2ver}) tend to the same value,
i.e. $D\ln\left(I^{\frac{1}{2}}\right)$, and therefore also the right hand sides can be set
to be equal in this limit. Moreover, the ratio $\frac{\Psi_-}{\Psi_+}\rightarrow
-i\tan\left(\frac{2\pi k}{3}\right)$ in the same limit. Putting this all together,
and subtracting Eq.~(\ref{eqn:bianchimodD22}) from Eq.~(\ref{eqn:bianchimodD42}) 
we find that the following
relation between spin coefficients holds in the Petrov type D limit

\begin{equation}
\label{eqn:spincoeffrel}
\left(\rho+2\tilde{\epsilon}\right)\sin\left(\frac{4\pi k}{3}\right)+i\tilde{\lambda}\cos\left(\frac{4\pi k}{3}\right)=0.
\end{equation}
Eq.~(\ref{eqn:spincoeffrel}) is valid for all three transverse frames, depending on
the value of $k$. If we assume to be in the
transverse frame that is also a quasi-Kinnersley frame, which corresponds to having
$k=0$, Eq.~(\ref{eqn:spincoeffrel}) reduces to $\tilde{\lambda}=0$, consistently with the
Goldberg-Sachs theorem. One can then use Eq.~(\ref{eqn:bianchimodD4}) 
to find the expression for $\rho$, obtaining
$\rho=D\ln I^\frac{1}{6}$, but the key point here is that in the quasi-Kinnersley frame
the Bianchi identities leave the expression for $\tilde{\epsilon}$ completely unresolved. 
However, it is really the expression for $\tilde{\epsilon}$ we are interested in, as it is the one
related to the spin-boost transformation. To obtain additional information on this
spin coefficient, we therefore analyze the Ricci identities.

\section{The Ricci identities}
\label{sec:ricci}

In this section, we will use the Ricci identities to understand how the 
spin coefficients $\epsilon$, $\gamma$, $\alpha$ and $\beta$ 
relate to the spin-boost parameter $\mathcal{B}$. We will first show that they
can be expressed as directional derivatives of the same function, and then determine
the equation that this function must satisfy in the limit of Petrov type D.

\subsection{Spin coefficients as directional derivatives}

We assume to be in the Petrov type D limit, where $\Psi_0=\Psi_1=\Psi_3=\Psi_4=0$
and also, as a consequence of the Goldberg-Sachs theorem, the four spin coefficients
$\lambda$, $\sigma$, $\nu$ and $\kappa$ are vanishing.
We begin with the following Ricci identity, obtained after adding and subtracting the product $\beta\epsilon$ on the right-hand side:

\begin{equation}
\label{eqn:firstricci}
D\beta-\delta\epsilon=\epsilon\left(\pi^*-\alpha^*-\beta\right)
+\beta\left(\rho^*+\epsilon-\epsilon^*\right).
\end{equation}
Introducing the rescaled spin coefficients, one can re-express this Ricci identity in terms
of $\tilde{\epsilon}$ and $\tilde{\beta}$, leading to 

\begin{equation}
\label{eqn:firstriccimod}
D\tilde{\beta}-\delta\tilde{\epsilon}=\tilde{\epsilon}\left(\pi^*-\alpha^*-\beta\right)
+\tilde{\beta}\left(\rho^*+\epsilon-\epsilon^*\right).
\end{equation}

Comparing Eq.~(\ref{eqn:firstricci}) with the expression of the commutator $\left[D,\delta\right]$ (again
assuming $\sigma=\kappa=0$)

\begin{equation}
\label{eqn:commudeltaD}
\left[D,\delta\right] = \left(\pi^*-\alpha^*-\beta\right)D
+ \left(\rho^*+\epsilon-\epsilon^*\right)\delta,
\end{equation}
it is possible to see that the Ricci identity is consistent with having $\tilde{\epsilon}=D\mathcal{H}_1$
and $\tilde{\beta}=\delta\mathcal{H}_1$, where $\mathcal{H}_1$ is a function to be determined. Using
the equivalent Ricci identity obtained after exchanging the tetrad vectors $\ell\leftrightarrow n$
and $m\leftrightarrow \bar{m}$

\begin{equation}
\label{eqn:firstriccireplace}
\Delta\tilde{\alpha}-\delta^*\tilde{\gamma}=\tilde{\alpha}\left(\gamma^*-\gamma-\mu^*\right)
+\tilde{\gamma}\left(\alpha+\beta^*-\tau^*\right),
\end{equation}
we obtain an equivalent result for the spin coefficients $\tilde{\gamma}$ and $\tilde{\alpha}$ and conclude
that they also can be expressed as $\tilde{\gamma}=\Delta\mathcal{H}_2$ and
$\tilde{\alpha}=\delta^*\mathcal{H}_2$. 

Using the properties of transformation of the spin coefficients  under the exchange operation
$\ell^{\mu} \leftrightarrow n^{\mu}$ and $m^{\mu} \leftrightarrow \bar{m}^{\mu}$, i.e.
$\tilde{\epsilon} \leftrightarrow -\tilde{\gamma}$ and
$\tilde{\alpha}\leftrightarrow- \tilde{\beta}$, we conclude that $\mathcal{H}_1=-\mathcal{H}_2=
\mathcal{H}$. The four spin coefficients can then be written as $\tilde{\epsilon}=D\mathcal{H}$,
$\tilde{\gamma}=-\Delta\mathcal{H}$, $\tilde{\beta}=\delta\mathcal{H}$ and 
$\tilde{\alpha}=-\delta^*\mathcal{H}$. 

The original spin coefficients are therefore given by

\begin{subequations}
\label{eqn:spinval}
\begin{eqnarray}
\epsilon&=& D\mathcal{H}-\frac{1}{2}D\ln\mathcal{B} = D\mathcal{H}_-, 
\label{eqn:limitepsilons} \\
\gamma&=& -\Delta\mathcal{H}-\frac{1}{2}\Delta\ln\mathcal{B}=-\Delta\mathcal{H}_+,
\label{eqn:spinlimitgammas} \\
\beta&=& \delta\mathcal{H}-\frac{1}{2}\delta\ln\mathcal{B} = \delta\mathcal{H}_-, 
\label{eqn:limitebeta} \\
\alpha&=& -\delta^*\mathcal{H}-\frac{1}{2}\delta^*\ln\mathcal{B}=-\delta^*\mathcal{H}_+,
\label{eqn:spinlimitalpha} 
\end{eqnarray}
\end{subequations}
where $\mathcal{H}_{\pm}=\mathcal{H}\pm\frac{1}{2}\ln\mathcal{B}$. We can now use
some of the remaining Ricci identities to find the expression for $\mathcal{H}$.

\subsection{The function $\mathcal{H}$}

We consider the two
following Ricci identities

\begin{subequations}
\label{eqn:finricciid}
\begin{eqnarray}
D\gamma-\Delta\epsilon&=&\alpha\left(\tau+\pi^*\right)+\beta\left(\tau^*+\pi\right) 
+\tau\pi 
\label{eqn:sixthricciin} \\
&-&\gamma\left(\epsilon+\epsilon^*\right)-\epsilon\left(\gamma+\gamma^*\right)+\Psi_2, \nonumber\\
\delta\alpha-\delta^*\beta&=&\mu\rho+\alpha\alpha^*+\beta\beta^*-2\alpha\beta 
\label{eqn:tenthricciin} \\
&+&\gamma\left(\rho-\rho^*\right)+\epsilon\left(\mu-\mu^*\right)-\Psi_2. \nonumber
\end{eqnarray}
\end{subequations}

The rescaled spin coefficients remove the spin-boost 
dependence in these identities, giving 

\begin{subequations}
\label{eqn:finricciidred}
\begin{eqnarray}
D\tilde{\gamma}-\Delta\tilde{\epsilon}&=&\tilde{\alpha}\left(\tau+\pi^*\right)+\tilde{\beta}\left(\tau^*+\pi\right) 
+\tau\pi 
\label{eqn:sixthricci} \\
&-&\tilde{\gamma}\left(\epsilon+\epsilon^*\right)-\tilde{\epsilon}\left(\gamma+\gamma^*\right)+\Psi_2, \nonumber\\
\delta\tilde{\alpha}-\delta^*\tilde{\beta}&=&\mu\rho+\tilde{\alpha}\alpha^*+\tilde{\beta}\beta^*-2\tilde{\alpha}\tilde{\beta} 
\label{eqn:tenthricci} \\
&+&\tilde{\gamma}\left(\rho-\rho^*\right)+\tilde{\epsilon}\left(\mu-\mu^*\right)-\Psi_2. \nonumber
\end{eqnarray}
\end{subequations}

Having just found that the reduced spin coefficients on the left-hand sides can be expressed
as directional derivatives of $\mathcal{H}$, one can
use the definition of double derivatives in the Newman-Penrose formalism to find an equivalent
form of Eq.~(\ref{eqn:finricciid}). In particular the equations we will make use of are 
the following

\begin{subequations}
\label{eqn:commide}
\begin{eqnarray}
D \Delta &=& -\left(\epsilon+\epsilon^*\right)\Delta+\pi\delta+\pi^*\delta^*+\ell^{\mu}n^{\nu}\nabla_{\mu}
\nabla_{\nu},  \label{eqn:ddeta}  \\
\Delta D &=& \left(\gamma+\gamma^*\right)D-\tau^*\delta-\tau\delta^*+n^{\mu}\ell^{\nu}\nabla_{\mu}
\nabla_{\nu},   \\
\delta \delta^* &=& \mu D -\rho^*\Delta-\left(\beta-\alpha^*\right)\delta^*+m^{\mu}\bar{m}^{\nu}\nabla_{\mu}\nabla_{\nu} ,  \\
\delta^* \delta &=& \mu^* D -\rho\Delta+\left(\alpha-\beta^*\right)\delta+\bar{m}^{\mu}m^{\nu}\nabla_{\mu}
\nabla_{\nu} .
\end{eqnarray}
\end{subequations}

As an example, we calculate the term $D\tilde{\gamma}$ on the left hand side of Eq.~(\ref{eqn:sixthricci}).
Using the property just found that in the Petrov type D limit $\tilde{\gamma}=-\Delta\mathcal{H}$, this term is given by
$D\tilde{\gamma}=-D\Delta\mathcal{H}$, and using Eq.~(\ref{eqn:ddeta})
this corresponds to

\begin{eqnarray}
-D \Delta \mathcal{H}&=& \left(\epsilon+\epsilon^*\right)\Delta\mathcal{H}- \pi\delta\mathcal{H}-\pi^*\delta^*\mathcal{H}\\&-&\ell^{\mu}n^{\nu}\nabla_{\mu}
\nabla_{\nu}\mathcal{H}. \nonumber
\end{eqnarray}
Substituting $\tilde{\alpha}=-\delta^*\mathcal{H}$ and $\tilde{\beta}=\delta\mathcal{H}$
gives

\begin{eqnarray}
D\tilde{\gamma}&=& -\left(\epsilon+\epsilon^*\right)\tilde{\gamma}- \pi\tilde{\beta}+\pi^*\tilde{\alpha}
-\ell^{\mu}n^{\nu}\nabla_{\mu}
\nabla_{\nu}\mathcal{H}.
\end{eqnarray}

Repeating the same procedure for $\Delta\tilde{\epsilon}$,
$\delta\tilde{\alpha}$ and $\delta^*\tilde{\beta}$, and comparing with the Ricci identities
in Eq.~(\ref{eqn:finricciidred}), one finds
the two following identities

\begin{subequations}
\label{eqn:accaeq}
\begin{eqnarray}
2\ell^{\mu}n^{\nu}
\nabla_{\mu}\nabla_{\nu}\mathcal{H}&=&
-2\pi\tilde{\beta}-2\tau\tilde{\alpha}-\pi\tau-\Psi_2, \label{eqn:acca1}\\
2m^{\mu}\bar{m}^{\nu}
\nabla_{\mu}\nabla_{\nu}\mathcal{H}&=&
-2\mu\tilde{\epsilon}-2\rho\tilde{\gamma}-\mu\rho+\Psi_2  \label{eqn:acca2}.
\end{eqnarray}
\end{subequations}
Subtracting Eq.~(\ref{eqn:acca2}) from Eq.~(\ref{eqn:acca1}), and using the expression
for the metric $g^{\mu\nu}=2\ell^{(\mu}n^{\nu)}-2m^{(\mu}\bar{m}^{\nu)}$,
 it is possible to obtain the final equation for $\mathcal{H}$:

\begin{equation}
\label{eqn:accafin}
\nabla^{\mu}\nabla_{\mu}\mathcal{H}+
\nabla^{\mu}\ln \left(I^{\frac{1}{6}}\right)\nabla_{\mu}\left(2\mathcal{H}+\ln I^{\frac{1}{12}}\right)=-2\Psi_2,
\end{equation}
where we have also used the fact that in the Petrov type D limit
$\rho=D\ln I^{\frac{1}{6}}$, $\mu=-\Delta\ln I^{\frac{1}{6}}$, $\tau=\delta\ln I^{\frac{1}{6}}$ and
$\pi=-\delta^*\ln I^{\frac{1}{6}}$.

In the next section we will solve Eq.~(\ref{eqn:accafin}) for the single black hole case to obtain
the condition on the spin-boost parameter.

\section{The Kerr limit}
\label{sec:kerr}

We can now apply the results we just found to the particular case
of the Kerr solution using Boyer-Lindquist coordinates.
The metric in this case reads

\begin{eqnarray}
ds^2&=&\left(1-\frac{2Mr}{\Sigma}\right)dt^2+\left(\frac{4Mar\sin^2\theta}
{\Sigma}\right)dtd\phi-\left(\frac{\Sigma}{\Gamma}\right)dr^2 \nonumber \\
&-&\Sigma d\theta^2-\sin^2\theta\left(\frac{r^2+a^2+2Mar\sin^2\theta}{\Sigma}
\right)d\phi^2,
\label{eqn:kerrboyer}
\end{eqnarray}
where $\Gamma=r^2-2Mr+a^2$ 
(in the usual notation this quantity is referred to as $\Delta$, but here, we
changed notation to avoid confusion with the derivative operator $\Delta$),
$\Sigma=r^2+a^2\cos^2\theta$, $M$ is the black hole mass and $a$ its
rotation parameter.

The Kinnersley tetrad in this coordinate system is given by

\begin{subequations}
\label{eqn:kerrtetrad}
\begin{eqnarray}
\ell^{\mu}&=&\left[\left(r^2+a^2\right)/\Gamma,1,0,a/\Gamma\right], 
\label{eqn:kerrtetradl} \\
n^{\mu}&=&\left[r^2+a^2,-\Gamma,0,a\right]/\left(2\Sigma\right), 
\label{eqn:kerrtetradn} \\
m^{\mu}&=&\left[ia\sin\theta,0,1,i/\sin\theta\right]/\sqrt{2}\bar{\rho},
\label{eqn:kerrtetradm}
\end{eqnarray}
\end{subequations}
where $\bar{\rho}=r+ia\cos\theta$. The solution 
for Eq.~(\ref{eqn:accafin}) in this particular coordinate system reads

\begin{equation}
\label{eqn:accasol}
\mathcal{H}=\frac{1}{2}\ln\left(\Gamma^{\frac{1}{2}}I^{\frac{1}{6}}\sin\theta \right).
\end{equation}

We now have all the elements to find the values of the spin coefficients $\epsilon$, $\gamma$,
$\beta$ and $\alpha$ in the limit of type D, and in particular the condition on the spin-boost
parameter. As already shown, the four spin coefficients $\epsilon$, $\gamma$, $\alpha$ and
$\beta$ can be written as follows

\begin{subequations}
\label{eqn:spinlimit}
\begin{eqnarray}
\epsilon&=& D\mathcal{H}-\frac{1}{2}D\ln\mathcal{B}, 
\label{eqn:limitepsilon} \\
\gamma&=& -\Delta\mathcal{H}-\frac{1}{2}\Delta\ln\mathcal{B},
\label{eqn:spinlimitgamma} \\
\beta&=& \delta\mathcal{H}-\frac{1}{2}\delta\ln\mathcal{B},
\label{eqn:spinlimitgbeta} \\
\alpha&=& -\delta^*\mathcal{H}-\frac{1}{2}\delta^*\ln\mathcal{B}.
\label{eqn:spinlimitalpha} 
\end{eqnarray}
\end{subequations}

This result
can be compared
with the expressions for the same spin coefficients in the Kinnersley tetrad, given by

\begin{subequations}
\label{eqn:spincoeffboostfin}
\begin{eqnarray}
\epsilon&=&0,
\label{eqn:epsilonlimit}\\
\gamma&=&\mu+\rho\rho^*\left(r-M\right)/2,
\label{eqn:gammalimit}\\
\beta&=&\cot\theta/(2\sqrt{2}\bar{\rho}),
\label{eqn:betalimit} \\
\alpha&=&\pi-\beta^*.
\label{eqn:alphalimit} 
\end{eqnarray}
\end{subequations}

Let us consider first the spin coefficient $\epsilon$. Using Eq.~(\ref{eqn:limitepsilon}) and the solution for $\mathcal{H}$ found in Eq.~(\ref{eqn:accasol}) we can
rewrite $\epsilon$ in the following way

\begin{equation}
\epsilon= \frac{1}{2}D\ln\left(\Gamma^{\frac{1}{2}}I^{\frac{1}{6}}\mathcal{B}^{-1}\sin\theta\right).
\label{eqn:epsdin}
\end{equation}
In order for this expression to be zero, the function inside the logarithm must be constant with
respect to the derivative operator $D$. Given the form of the Kinnersley tetrad
in Eq.~(\ref{eqn:kerrtetrad}), one concludes that the $D$ operator corresponds to the
simple $\partial_r$ derivative (assuming that the functions do not have a $t$ or $\phi$
dependence, which is indeed the case, as the Kerr spacetime is stationary and
axisymmetric). As a consequence of this, $\epsilon$ vanishes if the function on the
right hand side is a generic function only of  the coordinate $\theta$. This leads
to the following condition on the spin-boost parameter

\begin{equation}
\mathcal{B}=\mathcal{B}_0 f\left(\theta\right) I^{\frac{1}{6}} \Gamma^{\frac{1}{2}}\sin\theta,
\label{eqn:b1}
\end{equation}
where $\mathcal{B}_0$ is an integration constant. It can be easily shown that the
spin coefficient $\gamma$ given in Eq.~(\ref{eqn:gammalimit}) is consistent with Eq.~(\ref{eqn:b1}),
imposing no further condition on $f\left(\theta\right)$.

The 
spin coefficient $\beta$ can instead be used to find the unknown function
$f\left(\theta\right)$: 
the derivative operator $\delta$ is given by $\frac{1}{\sqrt{2}\bar{\rho}}\partial_{\theta}$
and is therefore related to the $\theta$ dependence of the spin-boost parameter.
A straightforward calculation gives $f\left(\theta\right)=\sin^{-1}\theta$,
consistent also with the spin coefficient $\alpha$. The final result for $\mathcal{B}$
reads

\begin{equation}
\mathcal{B}=\mathcal{B}_0 I^{\frac{1}{6}} \Gamma^{\frac{1}{2}},
\label{eqn:b12}
\end{equation}

When applied to the expression for the Weyl scalars given in Eq.~(\ref{eqn:psi2psi4qkt}), Eq.~(\ref{eqn:b12})
gives 

\begin{subequations}
\label{eqn:psi2psi4qktfinal}
\begin{eqnarray}
\Psi_0&=&\mathcal{B}_0^{-2}\cdot  \Gamma^{-1} I^{\frac{1}{6}}\left(\Theta- \Theta^{-1}\right),
\label{eqn:psi0qktf} \\
\Psi_2&=&-\frac{1}{2\sqrt{3}}\cdot I^{\frac{1}{2}}\left(\Theta+\Theta^{-1}\right),
\label{eqn:psi2qktf} \\
\Psi_4&=&\mathcal{B}_0^{2}\cdot  \Gamma I^{\frac{5}{6}}\left(\Theta- \Theta^{-1}\right).
\label{eqn:psi4qktf} 
\end{eqnarray}
\end{subequations}

It is remarkable how these expressions for the scalars immediately give the 
correct 
radial fall-offs at future null infinity once the peeling behavior of the Weyl tensor is 
assumed:
the function $\Gamma$ is only defined in the limit of Petrov type D and gives
no radial contribution at future null infinity; we find the same result for $\Theta$ as it is
the ratio of quantities that have the same radial behavior at future null
infinity. In conclusion, the 
quantities that give a contribution at future null infinity are the factors $I^{\frac{1}{6}}$,
$I^{\frac{1}{2}}$ and $I^{\frac{5}{6}}$; given that under the peeling assumption
$I \propto r^{-6}$, this corresponds to
$\Psi_0\propto r^{-1}$, $\Psi_2 \propto r^{-3}$ and $\Psi_4 \propto r^{-5}$.

The fact that we obtain radial fall-offs for $\Psi_0$ and $\Psi_4$ that are exchanged with
respect to the normal assumption of outgoing radiation 
(where $\Psi_0\propto r^{-5}$ and $\Psi_4 \propto r^{-1}$) is
not surprising: this is due to the fact that in the Kinnersley tetrad the null vector $\ell^{\mu}$ is
ingoing while $n^{\mu}$ is outgoing. The normal assumption requires instead the opposite 
situation where $\ell^{\mu}$ is outgoing and $n^{\mu}$ is ingoing. This means that one
needs to exchange $\ell^{\mu}\leftrightarrow n^{\mu}$ to have the right convention. This results in
$\mathcal{B}\rightarrow\mathcal{B}^{-1}$ and the Weyl scalars are changed to

\begin{subequations}
\label{eqn:psi2psi4qktfinal}
\begin{eqnarray}
\Psi_0&=&\mathcal{B}_0^{2}\cdot  \Gamma I^{\frac{5}{6}}\left(\Theta- \Theta^{-1}\right),
\label{eqn:psi0qktf} \\
\Psi_2&=&-\frac{1}{2\sqrt{3}}\cdot I^{\frac{1}{2}}\left(\Theta+\Theta^{-1}\right),
\label{eqn:psi2qktf} \\
\Psi_4&=&\mathcal{B}_0^{-2}\cdot  \Gamma^{-1} I^{\frac{1}{6}}\left(\Theta- \Theta^{-1}\right).
\label{eqn:psi4qktf} 
\end{eqnarray}
\end{subequations}
giving this time, as expected, the correct radial fall-offs for $\Psi_0$ and $\Psi_4$.

Eqs.~(\ref{eqn:psi2psi4qktfinal}) are the main result that we propose for wave extraction in numerical
relativity. As evident from the equations, the conditions on the spin coefficients do not completely fix
the values of the Weyl scalars, leaving the complex constant $\mathcal{B}_0$
undetermined. This is not surprising as such conditions involve
the directional derivatives along the tetrad null vectors and are therefore independent
of additional constant multiplication factors. The optimal value of this integration constant will have to be determined enforcing the values of the spin coefficients $\rho$, $\mu$, $\tau$ and $\pi$; the 
result of this calculation will be presented in a following numerical paper.
We are also investigating the comparison of these expressions with the analogous
quantities defined in the characteristic formulation of Einstein's equations
\cite{Lehner:2007ip,Gallo:2008pm,Gallo:2008sk}. As we expect,
this should give us more insights on how to choose this integration constant from a theoretical
point of view. This is the subject of future work on this topic.


\acknowledgments

The authors wish to thank Sarp Akcay, Emanuele Berti, Jeandrew Brink, Luisa T. Buchman, Robert Owen, Ulrich Sperhake and
Paul Walter 
for useful discussions and for careful proofreading of the manuscript. 
This work was supported by the DFG grant SFB/Transregio 7 ``Gravitational Wave Astronomy''.
OE acknowledges support from the Elitenetzwerk Bayern and LISA Germany.


\bibliographystyle{apsrev}
\bibliography{references}


\end{document}